# Beyond Dielectrics: Interfacial Water Polarization Governs Graphene-Based Electrochemical Interfaces

*Peiyao Wang[1,2], Gengping Jiang[3], Yuan Yan[1], Longbing Qu[2], Xiaoyang Du[2], Dan Li[2], and Jefferson Z. Liu[1]\**

[1]Department of Mechanical Engineering, The University of Melbourne, Parkville, VIC 3010, Australia

[2]Department of Chemical Engineering, The University of Melbourne, Parkville, VIC 3010, Australia

[3]Department of Applied Physics, College of Science, Wuhan University of Science and Technology, Wuhan, 430081, China

**ABSTRACT**

Water molecules are traditionally viewed as passive dielectric media in electrochemical systems. In this work, we challenge this conventional perspective through molecular dynamics simulations and theoretical analysis. We demonstrate that interfacial water is polarized distinctively from bulk water and excessively screens the electrostatic potential between ions and the surface, going beyond the classic electric double layer (EDL) model, which considered water merely a passive dielectric. This overscreening occurs because a significant portion of water polarization responds directly to the graphene surface, in addition to screening the electrostatic interaction between ions and charged surfaces. Furthermore, we reveal that this surface-induced interfacial water polarization governs the electric potential distribution, the EDL capacitance, and can even invert the electrode surface potential polarity, overriding the ion's contribution. These molecular-level insights underpin a revised EDL model that more accurately captures the electric and chemical potential distributions in the interfacial EDL regions.

**INTRODUCTION**

Aqueous electrolytes in contact with conductive solid materials are fundamental to numerous technologies related to energy, water, and biomedicine. Fundamental to the electrochemical interface is the formation of an electric double layer (EDL), where nearby liquid ions and water dipoles balance surface charges or mismatched work functions, leading to spatially reorganized electric and chemical potentials[1]. Numerous studies have shown that EDL-mediated potentials are crucial for various electrochemical and biological processes, including ionic pumps, charge storage, electrode stability, electrolyte conductivity, and electrochemical reaction kinetics[2-6]. Under the framework of the Gouy–Chapman–Stern (GCS) model, a foundation of EDL theory, the EDL consists of a near-surface compact layer (called Stern layer) and a diffuse layer behind it. Notably, under the operation of large electrode polarization and medium/high concentration electrolytes (> 0.1 M), which is common for many practical applications, the Stern layer dominates the electric and chemical potentials, governing the performance and efficiency of these technologies[7,8]. However, due to its sub-nanometric width and location buried between electrodes and the diffuse layer, the Stern layer has long remained elusive and is usually modeled as an effective physical parallel capacitor in the GCS model.

Recently, the investigations of EDL, particularly on the Stern layer, have seen promising progress on account of new spectroscopy techniques, such as optical-vibrational spectroscopies based on infrared absorption[9] and Raman scattering[10]. These state-of-the-art characterizations allow the possibility of probing the structural and electrical characteristics of ions and solvents within the Stern layer. The reported structural features of the water solvent suggest that its impact could be oversimplified within the GCS model, particularly at the Stern layer[7,8,11-15]. For example, based on graphene's well-defined atomically flat surface and its transparency in the relevant infrared frequency range, the surface-specific vibrational spectroscopy experiments, coupled with molecular simulations, revealed a significantly higher local density of water near the graphene surface compared to the bulk solution[16,17]. Additionally, the O−H bond of water in this region exhibits a directional preference pointing towards the graphene[12,13]. Given that water is a polar solvent, the recognized local density heterogeneity and specific molecular orientation would alter the spatial distribution of electric potentials beyond being a homogeneous medium assumed in the GCS model. Moreover, an asymmetric orientation response of interfacial water between positive and negative electric fields has also been reported[12-15], suggesting that water would not be a constant dielectric media as a function of electrode polarization. How these heterogeneous interfacial water structures will affect the electric potential spatial distribution across the EDL, particularly within the Stern layer, remains largely unclear.

In this work, we select graphene-interfaced aqueous electrolyte as a model system to scrutinize how interfacial water affects the local electric potential distributions and resultant electrochemical performance (Fig. 1a). Given that the local density and orientation of water dipoles would be altered by the graphene surface, ions, and charging states of graphene, we considered multiple systems to facilitate a controlled investigation, including graphene-interfaced pure water system

(Gr-water), graphene-interfaced NaCl aqueous electrolyte system (Gr-NaCl), and graphene-interfaced NaBF$_4$ aqueous electrolyte system (Gr-NaBF$_4$). NaCl is widely taken as the standard aqueous electrolyte. The BF$_4^-$ is considered because the interfacial water structure would be more susceptible to its strong non-electrostatic adsorption on the graphene surface[18,19]. We use a concentration of 0.8 M, typical for concentrated aqueous electrolytes[18,19], where interfacial and ion-specific effects are amplified, and sufficient sampling ensures statistical convergence in molecular dynamics (MD) simulations. Different charging states of graphene (electroneutral, positively charged, and negatively charged) are also investigated.

Given that experimental techniques often yield ensemble-averaged properties, and *ab initio* simulations are limited by scale, we employed classical MD simulations with well-validated force fields, enabling simulations over tens of nanoseconds—long enough to resolve collective and correlated water structuring that governs electrostatic behaviour at the interface. These simulations are then coupled with theoretical analysis to quantitatively dissect the contributions of water dipoles and ionic charges to the total electric potential. This combined approach provides both atomistic insight and predictive capability beyond conventional mean-field models (more simulation details in Supplemental Material sec.1-3 and our previous work[20,21]). Our approach thus captures both the microscopic and system-level physics. We reveal that the interfacial polarized water solvent overscreens the electrostatic potential between the ions and the surface, beyond being a passive dielectric medium. The interfacial water polarization dominates the electric potential spatial distribution and, thus, several key EDL properties. We extend the classic Poisson-Boltzmann EDL models to incorporate this interfacial water effect.

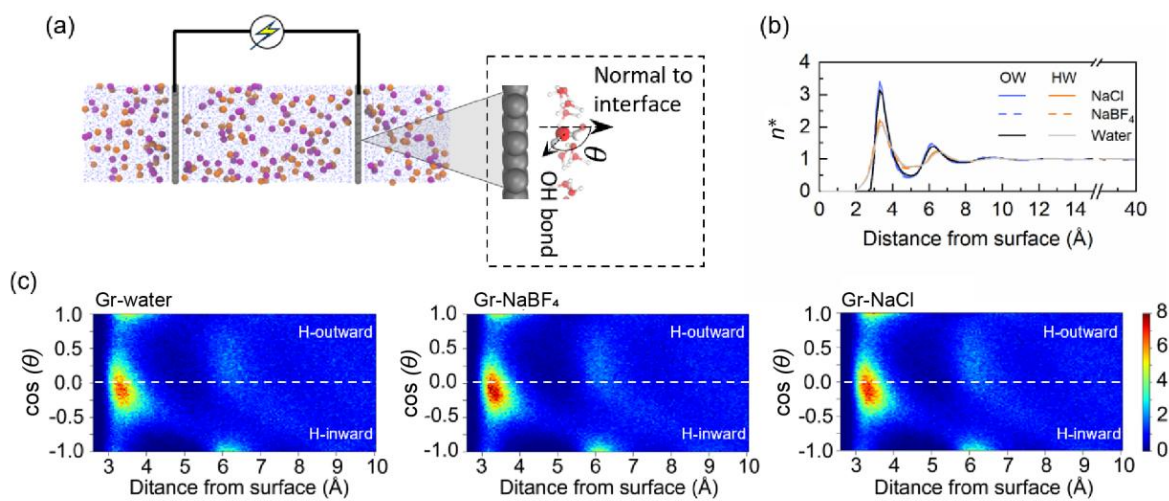

**Figure 1. The structured water at the uncharged interface.** (a) The representative model system with water molecules as points, cations as purple spheres, anions as orange spheres, and graphene as grey spheres. (b) Normalized local number distributions of water's oxygen (OW) and hydrogen (HW) atoms *vs.* a normal distance to the graphene surface. (c) Normalized orientation density of water's O−H bond as a normal distance to the graphene surface. The horizontal white line divides the map into bonds pointing away from the group (termed H-outward) and toward the surface (H-inward).

**RESULTS AND DISCUSSION**

**Crucial interfacial water in the electric potential distributions at uncharged graphene.** Our force field-based MD simulations reproduce the structural heterogeneity of water near graphene surfaces, in agreement with reported experimental characterizations[12-15] and *ab initio* MD results[17]. As reported previously, water molecules adjacent to uncharged graphene exhibit pronounced local density enhancement, showing a density peak ~3.5 times that of bulk water within 2-5 Å from the graphene surface (Fig. 1b). Also consistently[12-15], these interfacial water show preferential orientation of O−H bond towards the surface, where a portion of hydrogen atoms are positioned closer than the oxygen atoms to the graphene surface.

To further evaluate the interfacial water's orientation, we define an angle $\theta$ between the O−H bond and the surface normal (inset Fig. 1a). Consistent with previous findings[17], the two-dimensional (2D) orientation map reveals negative $\cos\theta$ values within 5 Å of the surface, confirming the prevalence of O−H bonds pointing towards the graphene (Fig. 1c). Importantly, we show that this orientation bias −and thus the interfacial polarization− persists across Gr-water, Gr-NaCl, and Gr-NaBF$_4$ systems, indicating that specific ion presence only marginally perturbs the interfacial water structure.

What distinguishes our work is the direct quantification of how this interfacial water polarization impacts the electric potential distribution across the EDL. Figure 2a-b compares the total electric potential $\phi_{\text{pzc,total}}$ in Gr-NaCl or Gr-NaBF$_4$ systems with that of pure Gr-water system ($\phi^0_{\text{pzc,H2O}}$). Strikingly, all three systems exhibit nearly identical near-surface potential oscillations: up to ~1.15 V peak-to-peak and decaying to zero beyond ~10 Å. These non-monotonic, oscillatory potential profiles contrast sharply with the monotonic decay predicted by classical GCS model and underscore the dominate role of polarized water in shaping the local electric field.

This influence is further reflected in the simulated potential of zero charge (PZC) (Fig. 2a-b and supplementary Table 1). The sign of PZC in Gr-NaCl and Gr-NaBF$_4$ systems shows an inconsistent trend with the ionic charge polarity at interfaces. For example, the Gr-NaCl system shows a PZC of 0.33 V. However, Na$^+$ and Cl$^-$ ions show a nearly equal local distribution at the interfacial region, which cannot account for such a significant positive PZC value (lower plot, Fig. 2a). It is even more striking in Gr-NaBF$_4$, which exhibits a positive PZC of 0.26 V (Fig. 2b and supplementary Table 1). This contrasts with the negative charge polarity of ions inferred from the spatial distribution of NaBF$_4$ ions, where the BF$_4^-$ ions show a closer adsorption distance and around three times higher local absorption density than Na$^+$ ions (lower plot, Fig. 2b). Note that our simulated PZC values are close to the reported experimental results[13].

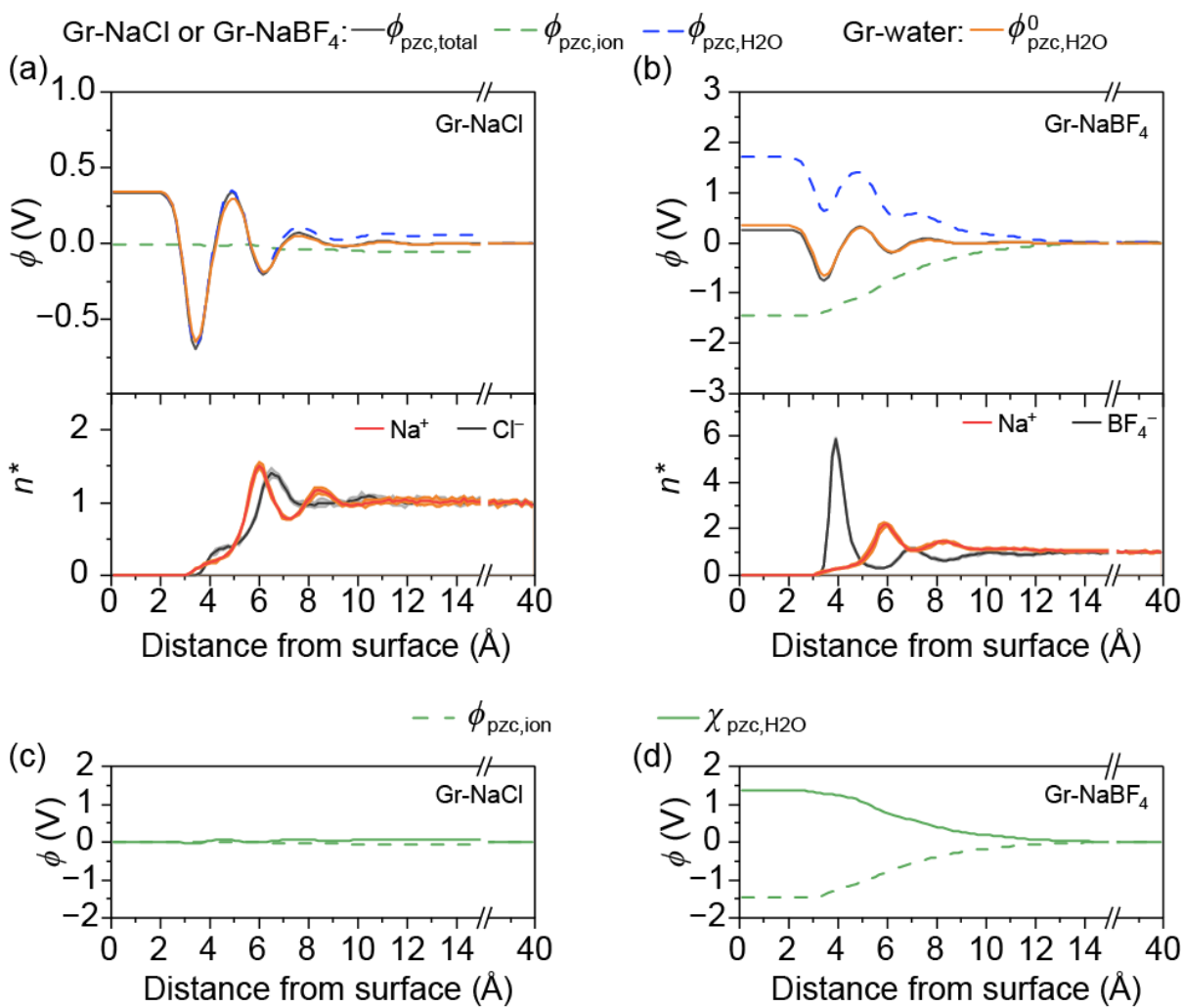

**Figure 2. Interfacial water polarization dominates the electrolyte electric potential at the uncharged graphene surface.** (a-b) The upper plots are the electric potential profiles, including $\phi_{pzc,total}$, $\phi_{pzc,ion}$, and $\phi_{pzc,H2O}$ in Gr-NaCl (a) and Gr-NaBF$_4$ (b). Additionally, the electric

potential profile in the Gr-water system, $\phi^0_{\text{pzc,H2O}}$, is included for a comparison. The lower plots are the normalized local number distributions of ions with respect to the normal distance from the uncharged graphene surface in the Gr-NaCl system (a) and Gr-NaBF$_4$ system (b). (c-d) The electric potential profiles, $\phi_{\text{pzc,ion}}$ and $\chi_{\text{H2O}}$, in Gr-NaCl (c) and Gr-NaBF$_4$ (d) systems, respectively.

**Interfacial water beyond a passive dielectric medium.** Given that the aforementioned $\phi_{\text{pzc,total}}$ in the Gr-NaCl or Gr-NaBF$_4$ system is dictated by water's contribution, on top of ion's influence, we carried out a further quantitative investigation on water's role by decomposing the $\phi_{\text{pzc,total}}$ into ion's contribution ($\phi_{\text{pzc,ion}}$) and water's contribution ($\phi_{\text{pzc,H2O}}$) based on the linearity of Maxwell's equations: $\phi_{\text{pzc,total}} = \phi_{\text{pzc,ion}} + \phi_{\text{pzc,H2O}}$. Specifically, $\phi_{\text{pzc,total}}$ and $\phi_{\text{pzc,H2O}}$ are determined by integrating the corresponding ion spatial charge and water polarization distributions, respectively (Supplementary Material Sec.3). By comparing the $\phi_{\text{pzc,ion}}$ and $\phi_{\text{pzc,H2O}}$ in either Gr-NaCl or Gr-NaBF$_4$ system, we discovered an apparently surprising overscreening effect: $\phi_{\text{pzc,H2O}}$ shows a larger magnitude over $\phi_{\text{pzc,ion}}$, particularly at the near-surface region (Fig. 2a-b and supplementary Table 1). For example, in Gr-NaCl, $\phi_{\text{pzc,ion}}$ is nearly flat as a function of normal distance from the graphene surface, and its PZC is −0.01 V. By contrast, $\phi_{\text{PZC,H2O}}$ shows a non-monotonic profile, and its PZC is 0.34 V. In Gr-NaBF$_4$, $\phi_{\text{pzc,ion}}$ shows a negative PZC of −1.45 V, while the $\phi_{\text{pzc,H2O}}$ shows a PZC of 1.71 V, overriding the magnitude of $\phi_{\text{pzc,ion}}$ and leading to a positive total PZC value.

The observed overscreening effect goes beyond the scope of the classic dielectric theory for aqueous electrolytes. In this conventional framework, water is modeled as a passive dielectric media that partially attenuates the electrostatic interaction among surface charge and electrolyte ions, which $\phi_{\text{pzc,H2O}}$ should be in opposite signs and smaller magnitudes than $\phi_{\text{pzc,ion}}$[22]. The critical interfacial water on overall electrical potentials and the puzzling overscreening effect indicate that in these ion-involved EDL systems, the water polarization, particularly those at the interfacial region, may not all serve to screen ions. Instead, a portion of water's polarization responds to the presence of a graphene surface. We dissect $\phi_{\text{pzc,H2O}}$ in Gr-NaCl and Gr-NaBF$_4$ systems into two components: (1) a quenched component for interfacial water polarization $\phi^0_{\text{pzc,H2O}}$; and (2) the residual perturbation component $\chi_{\text{H2O}}$ due to the presence of ions: $\phi_{\text{pzc,H2O}} = \phi^0_{\text{pzc,H2O}} + \chi_{\text{H2O}}$. The quenched component is adopted from the Gr-water system (orange lines in Fig. 2a-b).

Figure 2 c-d compares the profiles of $\chi_{\text{H2O}}$ and $\phi_{\text{pzc,ion}}$ in the uncharged Gr-NaCl and Gr-NaBF$_4$, respectively. In either case, the $\chi_{\text{H2O}}$ manifests as a nearly mirror image of $\phi_{\text{pzc,ion}}$ in terms of magnitude and local fluctuations, indicating that the perturbation $\chi_{\text{H2O}}$ term serves as a linear dielectric screening to $\phi_{\text{pzc,ion}}$. Indeed, the ratio of $\phi_{\text{pzc,ion}}/(\phi_{\text{pzc,ion}} + \chi_{\text{H2O}})$ in Fig. 2c-d is close to the dielectric constant of water in our MD simulations (Supplemental Material Sec. 8). This confirms our perspective that not all interfacial water polarization serves to screen ions (and thus the observed overscreening). Owing to the high value of water e$_r$, $\phi_{\text{pzc,ion}}$ and $\chi_{\text{H2O}}$ nearly compensate each other. As such, the interfacial water polarization, $\phi^0_{\text{pzc,H2O}}$, dominates the total electric potential (Fig. 2) $\phi_{\text{pzc,total}} = \phi_{\text{pzc,ion}} + \phi^0_{\text{pzc,H2O}} + \chi_{\text{H2O}} \approx \phi^0_{\text{pzc,H2O}}$.

We then evaluate the influence of interfacial water at charged graphene surfaces. The changes in electric potential profiles relative to the uncharged conditions are examined. A prefix $\Delta$ is used in the following to denote such changes, e.g., $\Delta\varphi_{\text{total}} = \varphi_{\text{ele,total}} - \varphi_{\text{pzc,total}}$ representing the total potential change. In Gr-NaCl and Gr-NaBF$_4$, $\Delta\varphi_{\text{total}} = \Delta\varphi^0_{\text{H2O}} + \Delta\varphi_{\text{ion}} + \Delta\chi_{\text{H2O}} = \Delta\varphi^0_{\text{H2O,I}} + \Delta\varphi^0_{\text{H2O,II}} + \Delta\varphi_{\text{ion}} + \Delta\chi_{\text{H2O}}$. Here, we further decompose $\Delta\varphi^0_{\text{H2O}}$ into the interface water part $\Delta\varphi^0_{H2O,II}$ and the rest contribution in bulk water region $\Delta\varphi^0_{H2O,I}$ of the Gr-water system. The bulk water shows linear dielectric response to the surface charge, *i.e.*, $\Delta\varphi^0_{H2O,I} = \frac{\Delta\phi_E(z)}{\varepsilon_r}$ in our MD simulations (Fig. S7).

In Fig. 3a-d, $\Delta\varphi_{total}$ is predominantly shaped by $\Delta\varphi^0_{H2O,II}$ for either positive or negative surface charges. The $(\Delta\chi_{H2O} + \Delta\varphi^0_{H2O,I})$ exhibit nearly mirror of $\Delta\phi_{ion}$, analogous to the zero surface charge case (Fig. 2). This is sensible because $\Delta\varphi^0_{H2O,I}$ shows a linear dielectric response to the surface charge and $\Delta\chi_{H2O}$ shows the additional linear dielectric response to ions. Again, owing to the high $e_r$ of water, they nearly cancel each other, leading to the observed dominance of $\Delta\varphi^0_{H2O,II}$ (*i.e.*, $\Delta\varphi^0_{H2O,II}/\Delta\varphi_{total} \sim$ 0.94 to 0.97 at surface for Gr-NaCl and Gr-NaBF$_4$). As a result, the calculated capacitance values in these three systems are very close (Supplemental Material Sec.6).

Our quantitative analysis on MD simulation-based electric potential profiles demonstrates a critical insight: not all water polarizations are to screen ions. Instead, a significant portion responds to the presence of a graphene surface. This interface water polarization dominates the overall electric potential profiles and, thus, several key electrochemical metrics, such as the PZC value, surface potential, and EDL capacitance.

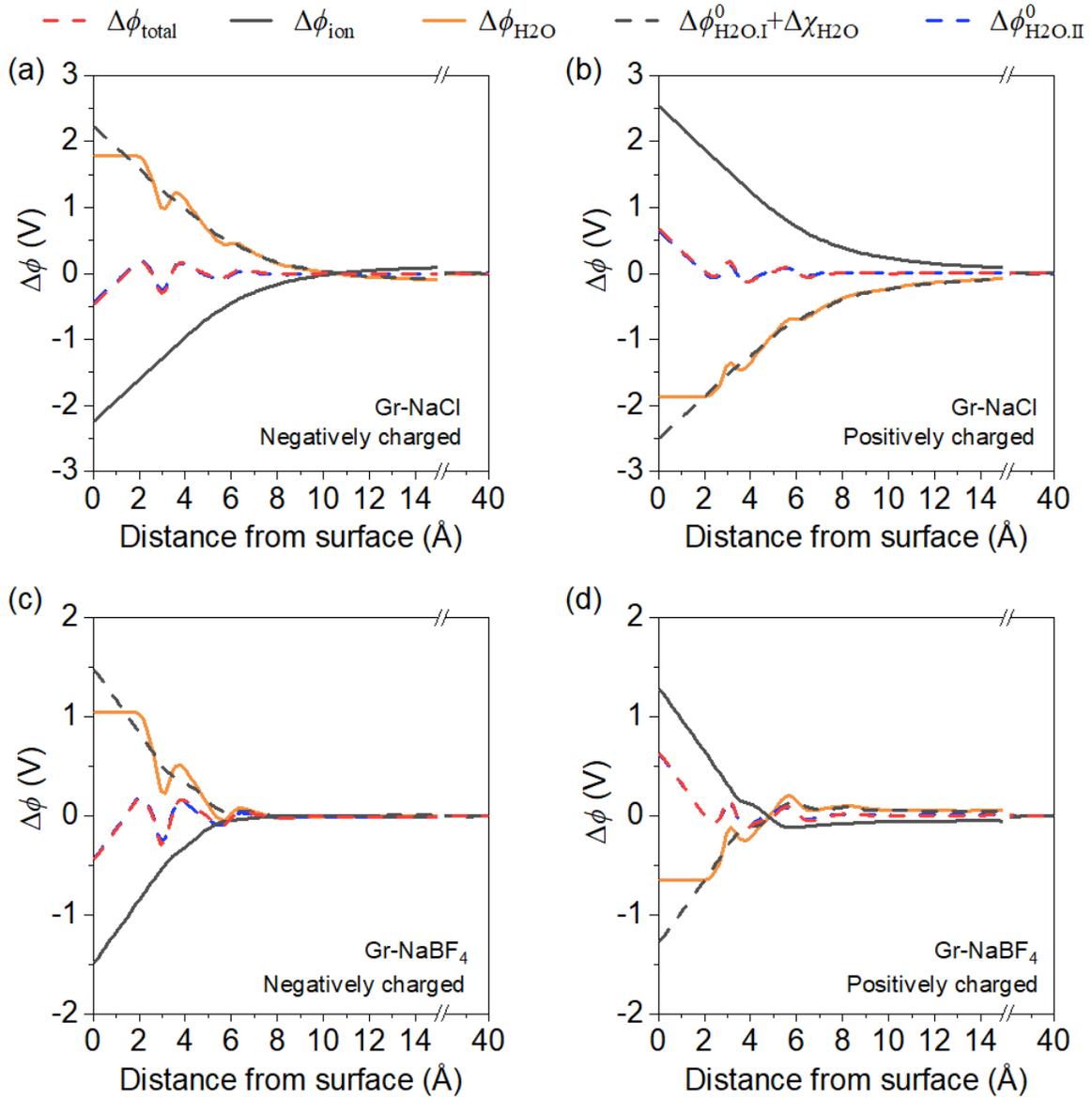

**Figure 3. The changes in the electric potential profiles at charged graphene surfaces relative to the uncharged case.** (a) and (b) compare the Gr-NaCl system with the Gr-water system: at a negatively charged surface (a) and at a positively charged surface (b). (c) and (d) compare the Gr-NaBF$_4$ system with the Gr-water system: at the negatively charged surface (c) and the positively charged surface (d). In each scenario, the sub-components of $\Delta\phi_{\text{total}}$, $\Delta\phi_{\text{ion}}$, ($\Delta\phi^0_{\text{H2O,I}}(z)$ + $\Delta\chi_{\text{H2O}}(z)$) and $\Delta\phi^0_{\text{H2O,II}}(z)$ are computed. The charge densities, $\sigma_s = \pm 0.00938$ $e$ C-atom$^{-1}$ (6.0

μC cm$^{-2}$), are adopted to model the charged graphene sheet, which is estimated based on the typical capacitance value of ~5-10 μF cm$^{-2}$ and an electric potential of ~1.0 to 1.2 V.

**The extended Poisson-Boltzmann EDL models.** The classic Poisson-Boltzmann (PB) equation models the screening effect of water solvent on ions and charged surfaces, where a linear dielectric parameter is adopted. Our above analysis shows that it effectively ignored the interface water polarization $\phi^0_{\text{pzc,H2O}}$ and $\Delta\varphi^0_{\text{H2O,II}}$. This motivated us to extend the PB equation for an advanced EDL model to mitigate this gap.

For the uncharged system scenario, $\phi_{\text{pzc,total}}$ can be decomposed into three parts:

$$\phi_{\text{pzc,total}}(z) = \underbrace{\phi^0_{\text{pzc,H2O}}(z) + \chi_{\text{H2O}}(z)}_{\phi_{\text{pzc,H2O}}} + \phi_{\text{pzc,ion}}(z) \tag{1}$$

Based on the observed linear dielectric screening effect of $\chi_{\text{H2O}}(z)$ on $\phi_{\text{pzc,ion}}(z)$ in MD simulations, the classic Poisson equation can be adopted,

$$\nabla^2(\phi_{\text{pzc,total}}(z) - \phi^0_{\text{pzc,H2O}}(z)) = -\frac{\rho_{\text{ion}}(z)}{\varepsilon_r \varepsilon_0} \tag{2}$$

where $\varepsilon_r$ is the relative dielectric constant, $\varepsilon_0$ is the dielectric constant at vacuum, and $\rho_{\text{ion}}(z)$ is the ionic local charge density. The $\rho_{\text{ion}}(z)$ can be expressed by the Boltzmann distribution:

$$\rho_{\text{ion}}(z) = \sum e z_i c_i = e c_i^0 \sum z_i e^{\frac{-[ez_i(\phi_{\text{pzc,total}}(z) - \phi^0_{\text{pzc,H2O}}(z)) + V_i^{\text{PMF}}]}{k_B T}} \tag{3}$$

where $c_i^0$ is the bulk concentration of ion species $i$, $k_B$ is the Boltzmann constant, and $T$ is temperature. The energetic term in the Boltzmann distribution should embody the total electric potential $\phi_{\text{pzc,total}}(z)$ as well as the non-electrostatic contributions like van der Waals interaction, steric effect, and solvation effect. Here we adopt the potential of mean force (PMF) $V_i^{\text{PMF}}$ determined in our MD simulations (Fig. S5) to describe the non-electrostatic contribution. $V_i^{\text{PMF}}$ describes the free energy variation of a single ion approaching an uncharged graphene surface along its normal direction. Note that the MD computed $V_i^{\text{PMF}}$ includes the interfacial water polarisation electric potential $\phi_{\text{pzc,H2O}}^0(z)$. To avoid double counting, it is excluded in Eq. (3). Equations (2) and (3) form our extended PB equation for the uncharged scenario (details in Supplemental Material).

For the charged system scenarios (denoted using subscript *ele*), the overall electric potential of the system can be decomposed as:

$$\phi_{\text{ele,total}}(z) = \phi_{\text{pzc,total}}(z) + \Delta\phi_{\text{total}}(z) \qquad (4)$$

$$= \phi_{\text{pzc,H2O}}^0(z) + \chi_{\text{H2O}}(z) + \phi_{\text{pzc,ion}}(z) + \Delta\phi_{\text{H2O}}^0(z) + \Delta\chi_{\text{H2O}}(z) + \Delta\phi_{\text{ion}}(z)$$

$$= \phi_{\text{pzc,H2O}}^0(z) + \Delta\phi_{\text{H2O,II}}^0(z) + \chi_{\text{H2O}}(z) + \phi_{\text{pzc,ion}}(z) + \Delta\phi_{\text{H2O,I}}^0(z) + \Delta\chi_{\text{H2O}}(z) + \Delta\phi_{\text{ion}}(z)$$

Based on the observed linear dielectric behaviors of water to surface charge and ion in Fig. 1 and 2 for the last five terms in Eq. (4), we can derive the Poisson equation to model the relation between ion distribution and the electric potential:

$$\nabla^2(\phi_{\text{ele,total}}(z) - \phi^0_{\text{pzc,H2O}}(z) - \Delta\phi^0_{\text{H2O,II}}(z)) = -\frac{\rho_{\text{ion}}(z)}{\varepsilon_r \varepsilon_0} \quad (5)$$

The local ion charge density $\rho_{\text{ion}}(z)$ is again described via the Boltzmann distribution:

$$\rho_{\text{ion}}(z) = \sum ez_i c_i = ec_i^0 \sum z_i e^{\frac{-[ez_i(\phi_{\text{ele,total}}(z) - \phi^0_{\text{pzc,H2O}}(z) - \Delta\phi^0_{\text{H2O,II}}(z)) + V_i^{\text{PMF}}]}{k_B T}} \quad (6)$$

The energetic term including the total electrical potential $\phi_{\text{ele,total}}(z)$ and the non-electrostatic interactions. We model the non-electrostatic term as $V_i^{\text{PMF}} - ez_i\phi^0_{\text{pzc,H2O}}(z) - ez_i\Delta\phi^0_{\text{H2O,II}}(z))$, where the first two terms are adopted from the zero surface charge case (Eq. (3)) and the last term is used to approximate the surface charge induced variation. This approximation is motivated by insights from Becker et al[23]. They found that the ion distribution seems not influenced by the strong electric field from the surface water polarization $\Delta\phi^0_{\text{H2O,II}}(z)$, which was partly attributed to the ion hydration shell structure change. It is likely that the non-electrostatic interaction variation resulting from the surface charge could compensate $\Delta\phi^0_{\text{H2O,II}}(z)$ and it is thus removed in Eq. (6). Our results (discuss later) demonstrate this approximation works well. More work should be done in future for this non-electrostatic interaction. The equations (5) and (6) form the modified PB equations to describe the electric potentials in charged scenarios. More details on model development are presented in Supplemental Material, including the boundary conditions.

Figure 4 and Supplemental Material Sec. 9-10 show that our model improved the prediction of the local electric potential distributions, compared to the classic PB equation (without Stern layer) and the GCS model (with Stern layer). Taking the Gr-NaCl system as an example (Fig. 4), the PB and GCS models predict the PZC to be 0 V, while our model predicted PZC value as 0.34 V, close to the MD result of 0.33 V. Additionally, the developed model predicts 0.99 V at the positively charged surface and −0.10 V at the negatively charged surface, closely matching MD results of 0.98 V and −0.12 V, respectively. In contrast, the PB equation, which adopts a uniform dielectric constant across the whole solution region, predicts only 0.01 V and −0.01 V, respectively (Fig. 4a). It is one magnitude smaller than the MD results. Compared to the PB equation, the GCS model shows some improvement, but significant differences still exist. For ion density distribution profiles, our models also show good consistency with MD simulation results (Supplemental Material Sec.10). The same improvement is also shown in the Gr-NaBF$_4$ system (Supplemental Material Sec.9). To further examine the approximated non-electrostatic term in Eq. (6), we compared the solved $\phi_{\text{ele,total}}(z) - \phi_{\text{pzc,H2O}}^0(z) - \Delta\phi_{\text{H2O,II}}^0(z)$ term from our model with direct MD results and obtained good agreement. These observed good agreements underscore the critical contribution of interfacial water polarization modulated by the graphene surface (i.e., $\phi_{\text{pzc,H2O}}^0$ at uncharged surface and $\Delta\phi_{\text{H2O,II}}^0$ on charged surfaces).

Note that in our extended PB model (Eq. (2-3) and (5-6)), the ion distribution $\rho_{\text{ion}}(z)$ and the net electric potential term without surface water contribution, i.e., $\phi_{\text{pzc,total}}(z) - \phi_{\text{pzc,H2O}}^0(z)$ or $\phi_{\text{ele,total}}(z) - \phi_{\text{pzc,H2O}}^0(z) - \Delta\phi_{\text{H2O,II}}^0(z)$, follow the framework of classic PB and GCS models. This could be the reason that the PB and GCS models can successfully predict many electrochemical experiments qualitatively or semi-quantitatively. After solving the PB equations and adding PZC potential and the surface water polarization to the obtained net electric potential, we can recover the full electrical potentials.

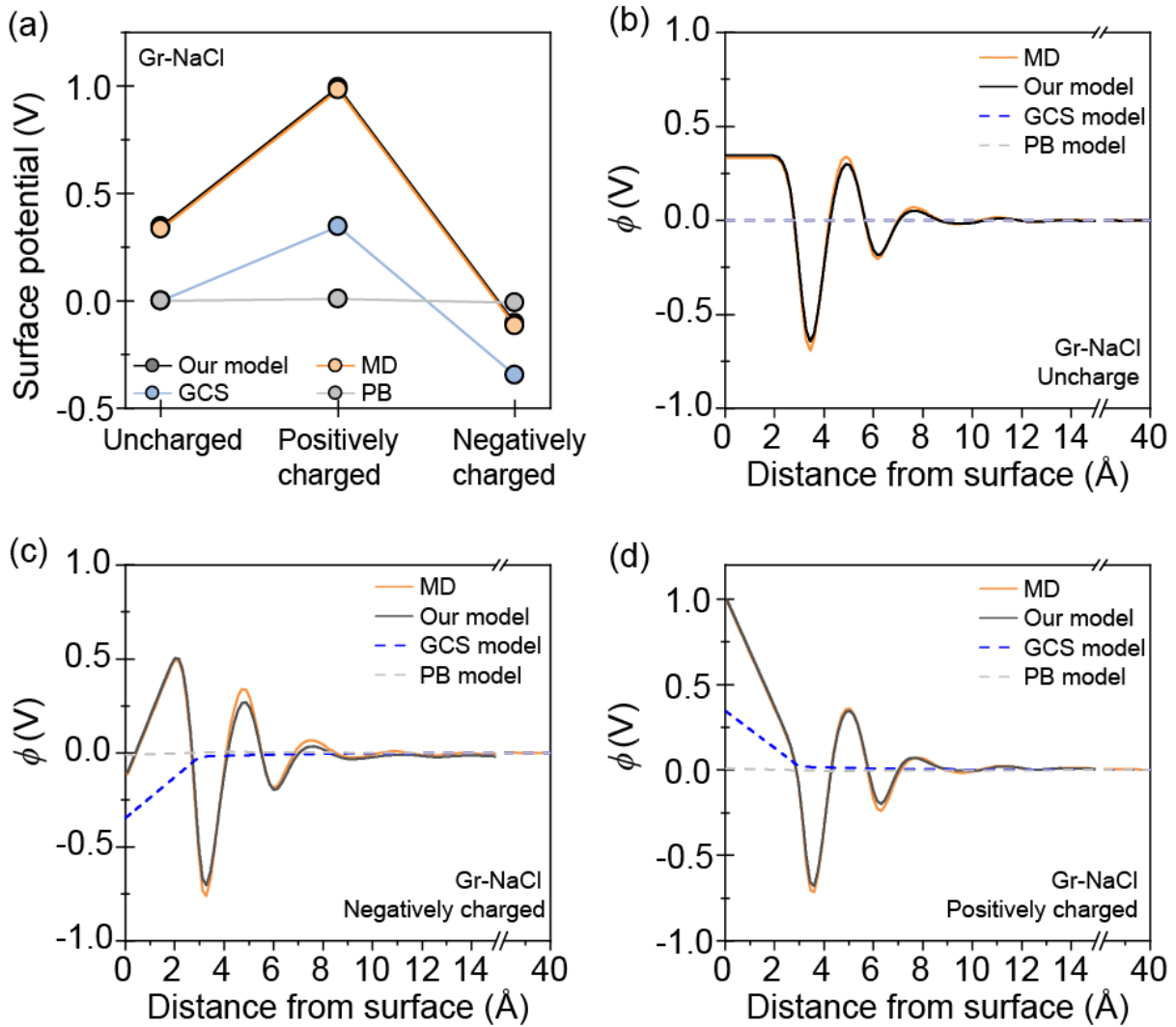

**Figure 4**. **Comparison of theoretical models in the Gr-NaCl system.** (a) The surface potential values predicted by different models, compared with the MD results. (b-d) The electric potential profile predicted by different models, compared with the MD results. Graphene surfaces are uncharged (a), negatively charged (b), and positively charged (c), respectively.

**Conclusions**

Our study reveals, through classical molecular dynamics and theoretical analysis, that interfacial water structure—rather than ionic distributions alone—plays a central role in governing electrostatic properties at graphene interfaces, both in uncharged and charged states. We show that the oriented, partially polarized water near the surface gives rise to pronounced oscillations in the electric potential, modulates the potential of zero charge (PZC), and dominates the EDL capacitance—challenging conventional predictions from classical mean-field models. These insights support the development of an extended Poisson–Boltzmann framework that incorporates orientational water effects, enabling improved predictions of local potentials, surface charges, and interfacial capacitance. Our findings advance the fundamental understanding and design of electrochemical interfaces relevant to a wide range of technologies, including supercapacitors, nanoionics, and neuromorphic iontronics.

The scope of our work focuses on systems with weakly polarizable surfaces, where electrostatics are primarily dictated by interfacial molecular organization and dipolar interactions. While our use of non-polarizable force fields does not capture induced electronic polarization in graphene, ions, or solvent, the strong agreement with experimental PZC values and *ab initio* results highlights the relevance and robustness of this approach in the targeted regime. Looking ahead, extending this framework to include explicit polarization or electronic-structure-based methods will be valuable for exploring systems with strong field effects, high surface charge, or metallic interfaces. We hope these findings inspire further efforts to bridge molecular-scale insights with continuum theories in complex electrochemical environments.

**Supporting Information**

Methodology of MD simulations; Methods for computing atom's local number distribution; Calculation of electric potential profiles; Comprehensive analysis of water orientation and local density distribution at graphene surfaces; Summary of capacitance values across different systems; Details of the developed EDL models; Comparison of theoretical model results in the Gr-NaBF4 system; Ion spatial distribution predicted by the developed EDL model.

**Corresponding Author**

*Jefferson Z. Liu: Department of Mechanical Engineering, The University of Melbourne, Parkville, VIC 3010, Australia; Email: zhe.liu@unimelb.edu.au

**Present Addresses**

Peiyao Wang — Max Planck Institute for Polymer Research, Mainz, Germany.

Dan Li — Department of Chemical and Biological Engineering, The Hong Kong University of Science and Technology, Clear Water Bay, Kowloon, Hong Kong 999077, China.

**Author Contributions**

All authors have given approval to the final version of the manuscript.

**ACKNOWLEDGMENT**

Jefferson Z. Liu acknowledges the funding from Australia Research Council (DP210103888, FT220100149). The molecular dynamics simulations were undertaken with the assistance of resources and services from the National Computational Infra-structure (NCI), which is supported by the Australian Government.

Table of Contents (TOC) graphics

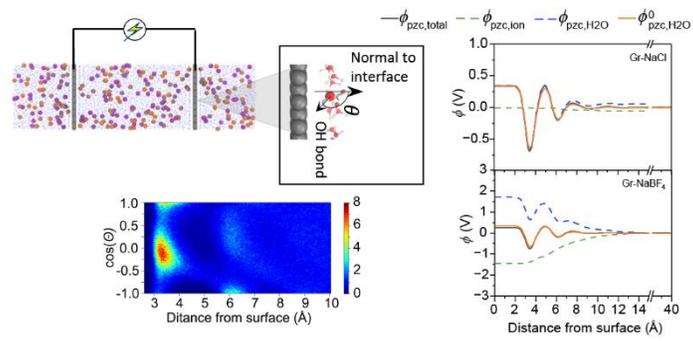